\documentclass[12pt,preprint]{aastex}

\slugcomment{Accepted by {\it the Astrophysical Journal Letters}}

\def\la{\mathrel{\hbox{\rlap{\hbox{\lower4pt\hbox{$\sim$}}}\hbox{$<$}}}}
\def\ga{\mathrel{\hbox{\rlap{\hbox{\lower4pt\hbox{$\sim$}}}\hbox{$>$}}}}

\shortauthors{Park}
\shorttitle{SNR 1987A}

\begin{document}

\title{Supernova Remnant 1987A: Opening the Future by Reaching the Past}

\author{Sangwook Park\altaffilmark{1}, Svetozar A. Zhekov\altaffilmark{2,3}, 
David N. Burrows\altaffilmark{1}, and  Richard McCray\altaffilmark{2}} 

\altaffiltext{1}{Department of Astronomy and Astrophysics, Pennsylvania State
University, 525 Davey Laboratory, University Park, PA. 16802; park@astro.psu.edu,
dburrows@astro.psu.edu}
\altaffiltext{2}{JILA, University of Colorado, Box 440, Boulder, CO. 80309; 
zhekov@super.colorado.edu, dick@jila.colorado.edu}
\altaffiltext{3}{On leave from the Space Research Institute, Sofia, Bulgaria}

\begin{abstract}

We report an up-turn in the soft X-ray light curve of supernova remnant (SNR)
1987A in late 2003 ($\sim$6200 days after the explosion), as observed 
with the {\it Chandra X-ray Observatory}. Since early 2004, the rapid increase of the 
0.5$-$2 keV band X-ray light curve can no longer be described by the exponential 
density distribution model with which we successfully fitted the data between 
1990 and 2003. Around day $\sim$6200, we also find that the fractional 
contribution to the observed soft X-ray flux from the decelerated shock 
begins to exceed that of the fast shock and that the X-ray brightening becomes 
``global'' rather than ``spotty''. We interpret these results as evidence that 
the blast wave has reached the main body of the dense circumstellar material all 
around the inner ring. This interpretation is supported by other recent
observations, including a deceleration of the radial expansion of the X-ray remnant, 
a significant up-turn in the mid-IR intensities, and the prevalence of the optical 
hot spots around the entire inner ring, all of which occur at around day 6000.  
In contrast to the soft X-ray light curve, the hard band (3$-$10 keV) X-ray light 
curve increases at a much lower  rate which is rather similar to the radio light curve. 
The hard X-ray emission may thus originate from the reverse shock 
where the radio emission is likely produced. Alternatively, the low increase rate 
of the hard X-rays may simply be a result of the continuous softening of the 
overall X-ray spectrum. 

\end{abstract}


\section {\label {sec:intro} INTRODUCTION}

The inner ring of supernova (SN) 1987A is believed to be a relic structure 
produced by the massive progenitor's equatorial stellar winds. The optical and 
X-ray brightness of SNR 1987A were expected to increase dramatically when the 
blast wave shock front hit this high-density structure (Borkowski et al. 1997a; 
1997b). The soft X-ray light curve of SN~1987A has been monitored on a regular
basis since 1991, first by the {\it ROSAT} satellite \citep{hasinger96} and then 
by the {\it Chandra X-ray Observatory} (Burrows et al. 2000; Park et al. 2002;
Park et al. 2004, P04 hereafter). Up to day $\sim$3000 (since the SN), the 
soft X-ray emission from SNR 1987A was faint  and was linearly increasing 
\citep{bur00}. The first optical hot spot emerged in 1997 \citep{pun97}, 
and subsequent soft X-ray observations by the {\it Chandra X-ray
Observatory} (beginning in October 1999) found an up-turn in the soft X-ray 
light curve \citep{bur00}, which deviated from the earlier linear increase. 
Since then the soft X-ray light curve has brightened dramatically 
(Park et al. 2002; P04). The rapid brightening of the soft X-ray light curve 
over the first four years of {\it Chandra} observations has been successfully 
described with a simple model which assumes a constant velocity shock propagating
into an ambient medium whose density increases exponentially with radius 
(P04; Park et al. 2005).

The emergence of optical hot spots in the inner ring has been interpreted in 
terms of the interaction of the shock front with irregularities (i.e., protrusions 
or clumps of dense gas, possibly produced by instabilities) on the inner surface 
of the inner ring \citep{michael00}. In this picture, the decelerated forward 
shock entering the dense protrusions is also expected to produce significant 
soft X-ray emission, and indeed soft X-ray bright spots have emerged at these 
{\it same} locations. The fraction of the shock front interacting with high density 
gas should increase gradually through the inclusion of more and more protrusions
until the entire inner ring is finally engulfed. Recent optical images obtained 
with {\it the Hubble Space Telescope} ({\it HST}) have indeed revealed spectacular 
developments of the optically bright spots all around the entire inner ring by 
the end of 2003 (e.g., Plate 1 in McCray 2005). 

Given the evidence for bright X-ray emission from the hot spots, it is
somewhat surprising that the simple model of a blast wave interacting with an 
exponential density distribution has provided such a good fit to the soft X-ray 
light curve up to this point. We report here a significant up-turn of the soft 
X-ray light curve at day $\sim$6200, which cannot be fitted with the simple 
exponential density model.

\section{\label{sec:obs} OBSERVATIONS \& DATA REDUCTION}

We have now performed a total of eleven monitoring observations of SNR 1987A with 
the Advanced CCD Imaging Spectrometer (ACIS) on board {\it Chandra}. We reduced 
the data following the methods described in our previous papers \citep{bur00,park02}. 
In order to derive the X-ray fluxes, we performed spectral fits using the two 
component plane-parallel shock model of P04, which provides a useful approximation 
of the complex temperature/velocity structure of the blast wave. The detailed 
description of the observations, data reduction and the image/spectral analyses 
of SNR 1987A, in which we obtained the X-ray fluxes and images presented in this 
{\it Letter}, can be found elsewhere (Park et el. 2005 in preparation). We note 
here that we use an updated version of the {\tt XSPEC} shock model 
appropriate for the non-equilibrium ionization plasma\footnote{The unpublished 
version of the updated model has been provided by K. Borkowski.}. We also 
re-generated the ancillary response functions using the ACIS {\tt caldb v3.00} for 
all eleven observations in order to consistently correct for the quantum efficiency 
degradation of the ACIS over the $\sim$five year period of the observations. 
The updated X-ray fluxes for all eleven ACIS-S3 observations are presented in 
Table~\ref{tbl:tab1}. 

\section{\label{sec:lcs} Soft X-ray Light Curves}

The latest X-ray light curves of SNR 1987A are shown in Figure~\ref{fig:fig1}. 
As of 2005 July, the observed 0.5$-$2 keV band X-ray flux is $F_X$ $\sim$ 
19.4 $\times$ 10$^{-13}$ ergs cm$^{-2}$ s$^{-1}$ ($L_X$ $\sim$ 1.42 $\times$ 
10$^{36}$ ergs s$^{-1}$ at $d$ = 50 kpc, after correcting for $N_H$ = 
2.35 $\times$ 10$^{21}$ cm$^{-2}$), which is an order of magnitude brighter 
than it was in 2000 January and two orders of magnitude brighter than in 1992. 
Figure~\ref{fig:fig1} shows that the simple ``exponential'' model of P04 
(the short-dashed curve) adequately fits the observed soft X-ray flux only until 
day $\sim$6200. The observed soft X-ray flux at later times increases significantly 
faster than the model prediction and cannot be fitted by this model. 
It is perhaps because this simple model does not accurately describe the soft
X-ray emission from the blast wave shock interacting with the more complex density 
profile. The effect of this complex shock structure may have recently become 
significant, as indicated by the development of the optically bright spots all 
around the inner ring.

The forward shock is expected to encounter an extremely steep increase in the 
ambient density as it interacts with the dense inner ring, which will result in 
a substantial deceleration of the overall shock velocity and a significant 
increase in the soft X-ray emission. We thus construct a more sophisticated model 
of the light curve which considers the fast and decelerated shock components 
separately. This model is motivated by the picture proposed in Zhekov et al. (2005), 
which reconciles the slow velocities inferred from X-ray line profiles with the 
much faster velocities obtained from the expansion of the X-ray remnant: i.e., 
the long-term expansion is dominated by the rapid shock in the low-density
H{\small II} region while the line widths are dominated by the slower forward
shock in the dense protrusions. Following the basic arguments of P04, the observed 
X-ray flux can be expressed as $F_X$ $\propto$ $n^2_0V_0T^{-0.6}_0$ + 
$n^2_rV_rT^{-0.6}_r$, where $n_0$, $V_0$, and $T_0$ are density, volume, and the 
electron temperature for the less dense H{\small II} region, respectively, while 
$n_r$, $V_r$, and $T_r$ are those for the shocked portion of the dense protrusions. 
As the blast wave enters the dense clumps, $T_r = T_0(n_0/n_r)$ can be assumed 
for an electron-ion temperature equilibration. The total emitting volume is given 
by $V_t$ = $V_r$  + $V_0$, and $f \equiv V_r/V_t$ is the volume filling factor 
of the dense clumps. The H{\small II} region is expected to have roughly constant 
density inside the inner ring, and the blast wave radius is therefore a power law
with time \citep{chev82}. For the case of SN1987A, with $\rho$ $\propto$ 
$t^{-3}v^{-9}$, the radius of the forward shock can be written as $R = R_0 + 
R'(t/t_0 - 1)^{2\over3}$, where $R_0$ is the radius (in arcseconds) at which 
the blast wave started to interact with the H{\small II} region, $t_0$ is the 
time at $R = R_0$, and $R'$ is a coefficient related to the radial expansion 
rate of the shock. Given that the number of optical spots has increased rapidly 
in recent years, we assume that the protrusion filling factor $f$ increases 
exponentially with radius to a value of unity at the radius of the inner ring. For
simplicity, we assume $V_0$ and $V_r$ $\propto$ $R^3$. The observed soft X-ray 
flux can then be expressed as $$F_X(\tau) = F_0 R(\tau)^3({\tau} - 1)^{0.4}\{1 + 
[{\eta}^{2.6} - 1]\exp{(-{\frac{R_r - R(\tau)}{D}})}\},$$ where $\tau$ $\equiv$ 
$t\over{t_0}$, $D$ is the characteristic {\it scale height} of the exponential 
filling factor, $\eta$ $\equiv$ ${n_r}\over{n_0}$, and $R_r$ is the radius of 
the inner ring. Based on the previous observations (P04 and references therein), 
we adopt $R_0$ = 0$\farcs$6, $t_0$ $\sim$ 1200 days, $R_r$ $\sim$ 0$\farcs$83. 
If we assume an average shock velocity of $v$ $\sim$ 3000 km s$^{-1}$ for the 
last several years, these results imply $R'$ $\sim$ 0$\farcs$07. 

We fit the soft X-ray light curve using this model in terms of $F_0$, $\eta$, 
and $D$. The best-fit model, displayed in Figure~\ref{fig:fig1} (the solid curve),
gives $F_0$ = 0.2$\pm$0.02 $\times$ 10$^{-13}$ ergs cm$^{-2}$ s$^{-1}$, 
$\eta$ = 7.3$\pm$0.4, and $D$ = 0$\farcs$018$\pm$0$\farcs$001, with $\chi^2/{\nu}$ 
= 13.1/15. The recent up-turn in the soft X-ray flux appears to be the result of
a significant increase in the fraction of the shock front interacting with the 
irregular inner boundary of the dense inner ring. As might be expected, the 
protrusion component now dominates the flux (providing $\sim$96\% of the total
flux at day 6700), while providing only a small contribution at early times (e.g., 
$\sim$9\% of the total flux at day 3000, shortly before the first optical bright 
spot appeared). We note that the density ratio $\eta$ is considerably lower than 
previous estimates (P04 and references therein). We attribute the low density ratio 
to the simplified diagnostics of our model which can only treat the detailed density
structure around the inner ring as an ``average'' value. 

\section{\label{sec:image} Soft X-ray Images}

Given the recent up-turn of the soft X-ray light curve and the prevalence of the
optical hot spots around the entire inner ring, the X-ray images might also reveal 
evidence for the brightening of the entire SNR. We therefore examined the spatial 
distribution of the soft X-ray brightening at two epochs. In Figure~\ref{fig:fig2}, 
we present the 0.5$-$2 keV band intensity ratios of SNR 1987A between two pairs 
of observations. In each panel, the gray-scales are cut above the average value 
in order to emphasize flux increases. It is evident that the strong brightening 
occurred mostly in the northern rim and the southeastern bright spot 
between 2000 and 2002 (Figure~\ref{fig:fig2}a). By contrast, the intensity 
increase between 2002 and 2005 (after day $\sim$6000) is nearly ubiquitous 
(Figure~\ref{fig:fig2}b), suggesting that the entire ring is now
increasing in X-ray brightness.

\section{\label{sec:disc} Discussion}

Days $\sim$6000 $-$ 6200 seem to be an important milestone in the evolution 
of SNR 1987A. This is the point after which the observed soft X-ray light curve 
starts to deviate considerably from the model prediction which had successfully 
described the data obtained over the previous $\sim$13 years. It is 
remarkable that this soft X-ray up-turn is accompanied by several other notable 
observational events. (1)~The blast wave shock kinematics obtained with our 
recent {\it Chandra} gratings observations indicated significantly slower 
velocities ($v$ $\sim$ 300 $-$ 1700 km s$^{-1}$) for the X-ray emitting hot gas 
than shock velocities previously estimated by X-ray and radio images ($v$ $\sim$ 
3000 $-$ 4000 km s$^{-1}$)~\citep{zhekov05}. (2)~The fractional contribution to 
the total observed X-ray flux from the decelerated shock (based on the two-shock
spectral modeling, Park et al. 2005 in preparation) has steadily increased over 
the past five years, and became dominant after day $\sim$6200 (Figure~\ref{fig:fig3}). 
(3)~Until day $\sim$5800, the radial expansion rate of the X-ray remnant was $v$ 
$\sim$ 4000 km s$^{-1}$ (P04). The latest {\it Chandra} data show that the X-ray 
radial expansion rate decreases substantially to $v$ $\sim$ 1600 km s$^{-1}$ after 
day $\sim$6200 (Racusin et al. 2005, in preparation). (4)~The entire inner ring 
became dominated by the optical hot spots by day $\sim$6000 (e.g., Plate 1 in McCray 
2005). (5)~Recent mid-IR observations of SNR 1987A, most likely indicating dust 
emission in the inner ring, show a remarkable brightening since day $\sim$6000 
\citep{bouchet05}. The optical/IR observations, X-ray spectral results, X-ray 
expansion measurements, soft X-ray images, and soft X-ray light curves are all 
consistent with our picture that the shock is now interacting with dense gas all 
around the inner ring.

\begin{figure}[]
\figurenum{1}
\centerline{\includegraphics[angle=0,width=0.85\textwidth]{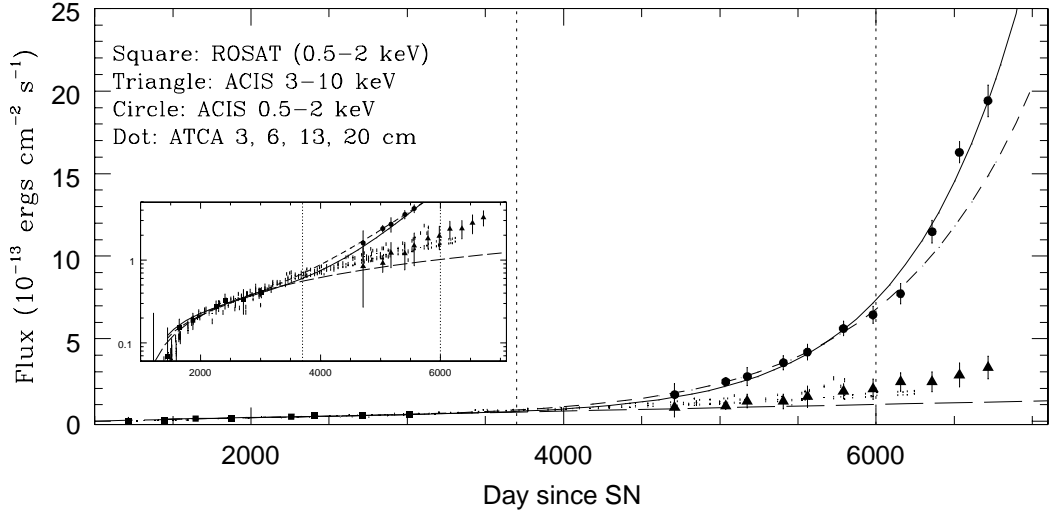}}
\figcaption[]{The composite light curves of SNR 1987A. The {\it ROSAT} data
are taken from Hasinger et al. (1996). The radio fluxes were obtained with the
{\it Australian Telescope Compact Array (ATCA)} (provided by L. Staveley-Smith)
and are arbitrarily scaled for the purpose of display. The solid curve is the 
best-fit model for the 0.5$-$2 keV band light curve ({\it ROSAT} + {\it Chandra}) 
as determined in this work (\S~\ref{sec:lcs}). The short-dashed curve is the best-fit 
model from P04, which is 
extrapolated after day 6000. The long-dashed curve is the linear model fitted
to the {\it ROSAT} data only and then extrapolated until day 7000. The two dotted 
vertical lines mark day 3700 and 6000, respectively. The inset is a log-scale 
presentation showing the details of the {\it ROSAT}, radio, and the hard X-ray 
light curves.   
\label{fig:fig1}}
\end{figure}

\begin{figure}[]
\figurenum{2}
\centerline{\includegraphics[angle=0,width=0.55\textwidth]{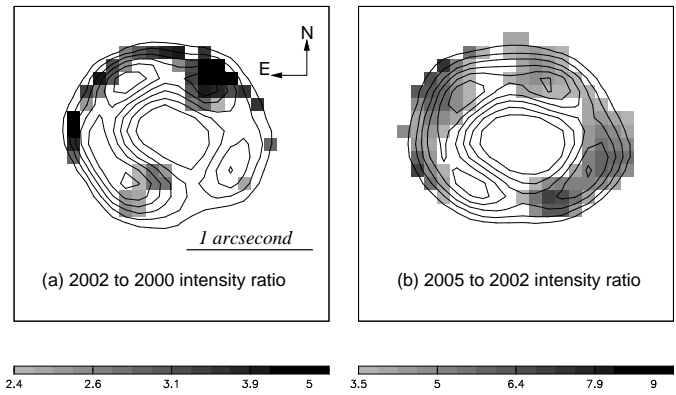}}
\figcaption[]{The 0.5$-$2 keV band intensity ratio of SNR 1987A. (a)
the 2002 December (day 5791) to 2000 December (day 5038) ratio, overlaid
with contours of the 0.5$-$2 keV image taken on 2002 December. (b)
the 2005 July (day 6716) to 2002 December (day 5791) ratio, overlaid with
contours of the 0.5$-$2 keV image taken on 2005 July. 
\label{fig:fig2}}
\end{figure}

\begin{figure}[]
\figurenum{3}
\centerline{\includegraphics[angle=0,width=0.45\textwidth]{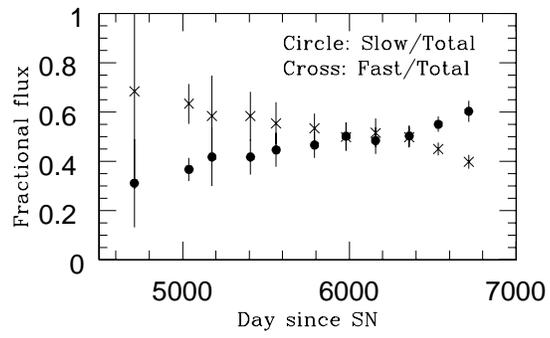}}
\figcaption[]{The fractional contributions of the slow and fast shock
components to the total 0.5$-$2.0 keV band flux, based on the spectral
fits of Park et al. 2005 (in preparation).
\label{fig:fig3}}
\end{figure}

\begin{figure}[]
\figurenum{4}
\centerline{\includegraphics[angle=0,width=0.55\textwidth]{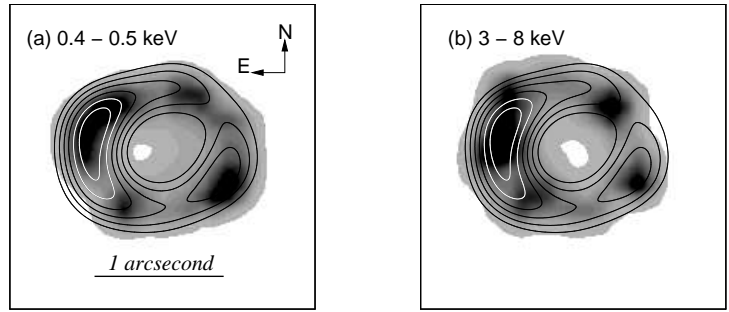}}
\figcaption[]{Gray-scale X-ray images of SNR 1987A (2005 July) overlaid with the
{\it ATCA} 9 GHz radio contours (2005 June). (a) The soft X-ray (0.4 $-$ 0.5 keV) 
vs. radio, (b) the hard X-ray (3 $-$ 8 keV) vs. radio. Both soft and hard X-ray
images contain similar photon statistics of $\sim$650 $-$ 750 counts. The radio 
images have been provided by B. Gaensler and L. Staveley-Smith. 
\label{fig:fig4}}
\end{figure}

On the other hand, the hard X-ray flux is increasing much less rapidly than the 
soft X-ray flux, although it is still steeper than the extrapolated {\it ROSAT} 
light curve (Figure~\ref{fig:fig1}, the long-dashed curve). We note that the hard 
X-ray light curve is similar to the radio light curves (Figure~\ref{fig:fig1}). 
The radio emission likely originates from  synchrotron emission from the
shocked ejecta behind the reverse shock \citep{man05}. The similarity between 
the light curves in the radio and the hard X-ray emission suggests that the 
hard X-rays might also be produced behind the reverse shock rather than the 
decelerated forward shock front. However, the radio map does not show a clearly
better correlation with the hard X-ray image than with the soft X-ray image
(Figure~\ref{fig:fig4}). Alternatively, the low rate of brightening in the hard 
X-ray light curve may simply result from the overall softening of the X-ray 
spectrum as an increasing fraction of the forward shock front decelerates. If 
the hard X-rays are truly originating from the same locations where the radio 
emission is produced, we may expect the hard X-ray and radio light curves to 
increase at the same rate in the future. Otherwise, the hard X-ray flux increase 
rate may continue to decline. As SNR 1987A continues to brighten, we may achieve 
sufficient photon statistics even in the hard X-ray band to distinguish between 
thermal and nonthermal nature, which will be critical to unveil the origin of 
the hard X-ray emission: e.g., broad (a few 10$^3$ km s$^{-1}$) emission lines 
may be detected if the hard X-rays originate from a thermal plasma behind the 
reverse shock. 

As the blast wave now sweeps through the inner ring, the soft X-ray brightening 
of SNR 1987A will be more spectacular than ever and the X-ray spectrum may 
significantly change in coming years. In this new stage, SNR 1987A will provide 
a unique, unprecedented opportunity for ``real time'' studies of the evolution 
of both the forward and reverse shocks. The soft X-ray emission from 
the main body of the inner ring would also be useful to study the past history 
of the progenitor star by monitoring the progress of the forward shock
through this material.  Continuous multi-wavelength monitoring of SNR 1987A, 
including in X-rays, will be essential for such studies.

\acknowledgments

The authors thank K. Borkowski for providing the unpublished version of the
shock model. We also thank B. Gaensler and L. Staveley-Smith for providing 
unpublished radio images and fluxes taken with {\it ATCA}. This work was 
supported in part by SAO under {\it Chandra} grant GO5-6073X.

\clearpage

\begin{deluxetable}{cccccc}
\tabletypesize{\footnotesize}
\tablecaption{X-ray flux of SNR 1987A.
\label{tbl:tab1}}
\tablewidth{0pt}
\tablehead{\colhead{Age\tablenotemark{a}} & \colhead{Observed Flux\tablenotemark{b} 
({\it ROSAT})} & \colhead{Age\tablenotemark{a}} & \colhead{Observed 
Flux\tablenotemark{b} ({\it Chandra})} & \colhead{Age\tablenotemark{a}} & 
\colhead{Observed Flux\tablenotemark{b} ({\it Chandra})} \\ 
\colhead{(days)} & \colhead{(0.5$-$2 keV)} & \colhead{(days)} & 
\colhead{(0.5$-$2 keV)} & \colhead{(days)} & \colhead{(3$-$10 keV)} }   
\startdata
1215 & $<$0.23 & 4711 & 1.61$\pm$0.66 & 4711 & 0.84$\pm$0.57 \\
1448 & 0.07$\pm$0.09 & 5038 & 2.40$\pm$0.22 & 5038 & 0.92$\pm$0.21 \\ 
1645 & 0.15$\pm$0.04 & 5176 & 2.71$\pm$0.54 & 5176 & 1.22$\pm$0.41 \\ 
1872 & 0.19$\pm$0.04 & 5407 & 3.55$\pm$0.43 & 5407 & 1.20$\pm$0.44 \\
2258 & 0.27$\pm$0.05 & 5561 & 4.19$\pm$0.46 & 5561 & 1.49$\pm$0.64 \\
2408 & 0.32$\pm$0.07 & 5791 & 5.62$\pm$0.45 & 5791 & 1.82$\pm$0.46 \\
2715 & 0.33$\pm$0.11 & 5980 & 6.44$\pm$0.52 & 5980 & 1.95$\pm$0.62 \\
3013 & 0.41$\pm$0.06 & 6157 & 7.73$\pm$0.62 & 6157 & 2.38$\pm$0.57 \\
 -   &   -           & 6359 & 11.48$\pm$0.69 & 6359 & 2.40$\pm$0.60 \\
 -   &   -           & 6533 & 16.29$\pm$0.65 & 6533 & 2.80$\pm$0.73 \\
 -   &   -           & 6716 & 19.41$\pm$0.97 & 6716 & 3.26$\pm$0.68 \\
\enddata

\tablenotetext{a}{Days since the SN explosion.}
\tablenotetext{b}{X-ray fluxes are in units of 10$^{-13}$ ergs cm$^{-2}$ s$^{-1}$.
The {\it ROSAT} fluxes are taken from Hasinger et al. (1996).}

\end{deluxetable}

\clearpage

\clearpage

\clearpage

\clearpage

\end{document}